# Determining sentiment in citation text and analyzing its impact on the proposed ranking index

Souvick Ghosh[1], Dipankar Das[1] and Tanmoy Chakraborty[2]

[1] Jadavpur University, Kolkata 700032, WB, India
{ souvick.gh, dipankar.dipnil2005}@gmail.com
[2] University of Maryland, College Park, MD 20742, USA
tanchak@umiacs.umd.edu

**Abstract.** Whenever human beings interact with each other, they exchange or express opinions, emotions and sentiments. These opinions can be expressed in text, speech or images. Analysis of these sentiments is one of the popular research areas of present day researchers. Sentiment analysis, also known as opinion mining tries to identify or classify these sentiments or opinions into two broad categories – positive and negative. Much work on sentiment analysis has been done on social media conversations, blog posts, newspaper articles and various narrative texts. However, when it came to identifying emotions from scientific papers, researchers used to face difficulties due to the implicit and hidden natures of opinions or emotions. As the citation instances are considered inherently positive in emotion, popular ranking and indexing paradigms often neglect the opinion present while citing. Therefore in the present paper, we deployed a system of citation sentiment analysis to achieve three major objectives. First, we identified sentiments in the citation text and assigned a score to each of the instances. We have used a supervised classifier for this purpose. Secondly, we have proposed a new index (we shall refer to it hereafter as M-index) which takes into account both the quantitative and qualitative factors while scoring a paper. Finally, we developed a ranking of research papers based on the M-index. We have also shown the impacts of M-index on the ranking of scientific papers.

## 1 INTRODUCTION

Sentiment analysis of citation contexts is an unexplored field in the area of sentiment analysis, primarily because of the existing myth that most of the research papers are cited positively in general. Furthermore, the negative citations are hardly explicit and the criticisms are often veiled. This lack of explicit sentiment expressions poses a major challenge for successful polarity identification. However, sentiment analysis of citations in scientific papers and articles is a new and interesting problem which can open up many exciting new applications in bibliographic search and bibliometrics [1].

There are many linguistic differences between scientific texts and other genres [2] and the scientific community has to undertake new approaches for correct classification. In our work, we have used various existing and novel features. We have considered n-grams, specialized science-specific lexical features, dependency relations, various word lists, sentiment lexicons and negation features for our research.

The importance of citation is due to the fact that it helps us in determining the impact of each cited paper. While most of the ranking indices rely solely on the number of citations each paper receives, we have added a qualitative measure to the ranking procedure. Our main goal is to use the sentiment information in each citation instance (qualitative) in addition to the number of citations (quantitative) to determine the worth of the paper. To the best of our knowledge, this is the first work that uses both the quantitative and qualitative indicators to determine the ranking of scientific papers.

This paper is structured in seven sections. First, we discuss the previous works that were done related to sentiment analysis of citations. Then, we explain how we prepared the corpus for this task. The next two sections explain the features used for classification and the classification procedure itself. The last three sections concentrate on the ranking algorithm, the results and the future work.

## 2    RELATED WORK

Automated citation sentiment analysis has emerged as a new research topic in natural language processing over the last decade [1], [15], [16]. An automated analysis would primarily take into account various linguistic cues, like tone of reference and any negative words and then make use of machine learning algorithms to evaluate the opinion of the author towards the cited paper.

Citation sentiment detection can also help researchers during search, by detecting problems with a particular approach. It can be used as a first step to scientific summarization [14], enable users to recognize unaddressed issues and possible gaps in the current research, and thus help them set their research directions [1]. Existing bibliometrics evaluation schemes, like H-index [8, 9], G-index [10], Impact Factor [11] and graph ranking algorithm like PageRank [12], focus mainly on the quantitative aspect of citations. However, for fair evaluation of a scientific paper, we need to consider the polarity of citation, or the qualitative aspect of citation. In many cases, we often find that the paper has been cited in a negative way, i.e., for the purpose of criticism. Bonzi et al. was one of the first proponents of this logic that if a cited work is criticized it should carry a lower or negative weight for bibliometric measures [13]. Abu-Jbara et al. worked on various linguistic analysis techniques for determining the purpose and polarity of citations [2].

Athar et al. [3] explored various sentence structure-based features for automatic identification of sentiment polarity in scientific literature. Athar et al. also worked on context-enhanced citation sentiment detection and studied the effectiveness of the length of the context window.

## 3    CORPUS PREPARATION

Most of the current work on citation sentiment detection focuses only on the citation sentences. The corpus has been obtained from Athar and groups [3]. It contains the source and the target paper, along with the citation sentence and its associated polarity (marked as 'o', 'p' or 'n' for *objective, positive* and *negative* respectively). This corpus has a total of 8736 sentences each of which is annotated manually with polarity. For our work, we selected 6736 instances for our training purposes and 2000 instances for our testing purposes.

Initially, we consider a baseline system with all the sentiments considered as neutral. As most of the citations are usually neutral (causing highly imbalanced classes), so the accuracy of the baseline system is quite well. It is also one of the reasons why we have to be careful while annotating them with a positive or negative tag as wrong tags might reduce the accuracy of the system to below baseline. We preprocessed the corpus to denote the polarity by three integers. + 1 for *positive*, -1 for *negative* and 0 for *neutral*. We also used a list of polar words and phrases which are specific to citation texts in order to identify the opinion of the citing author.

**Table 1.** Number of instances of each polarity in the dataset

|                | Positive | Neutral | Negative |
|----------------|----------|---------|----------|
| Entire dataset | 829      | 7627    | 280      |
| Training set   | 635      | 5888    | 213      |
| Test set       | 194      | 1739    | 67       |

## 4    FEATURE IDENTIFICATION

We evaluated the following features for identifying the sentiment polarity of the citation instances:

### 4.1    Automatic Sentiment (AS)

We have calculated an automatic sentiment score by splitting a sentence into a bag of words and then assigning score to each of the words. The words have been normalized before assigning scores. The score of individual words were formulated using SentiWordNet[1]. The sentiment score of the sentence is the sum of the scores of all the individual words multiplied by 100 (which helps in rounding the score). In the following example, the automatic score allocation by SentiWordNet is 43.0

*e.g.: Dasgupta and Ng (2007) improves over (Creutz, 2003) by suggesting a simpler approach. (Citing paper id 'W09-0805', cited paper id 'N07-1020')*

---

[1] http://sentiwordnet.isti.cnr.it/download.php

### 4.2 Positive polarity words (PPW)

We have used a list of words (unigrams, bigrams, trigrams and four-grams) with positive sentiment polarity. For each n-gram (up to n=4) present in the sentence we have compared it with the collection of positive n-grams to determine if there is a match. The most frequent unigrams, bigrams and trigrams which are specific to citation texts and positive in polarity are illustrated in tables 2-4. The frequencies were obtained from the training dataset. The number of trigrams and 4-grams are significantly less than that of unigrams and bigrams. Phrases like '*improve performance of*', '*very high accuracy*', '*most widely used*' and '*state of the art*' are generally used in conjunction with citation texts which denote positive polarity.

**Table 2.** Most frequent unigrams with positive polarity

| | | | |
|---|---|---|---|
| More (397) | Improvement (88) | Outperform (48) | Popular (35) |
| Most (308) | Important (86) | Correlate (47) | Efficient (31) |
| Improve (185) | High (82) | Higher (44) | Successful (30) |
| Best (153) | Effective (68) | Major (42) | Overcome (29) |
| Well (148) | Accurate (67) | Significant (39) | Consistent (23) |
| Better (141) | Development (67) | Highly (37) | Sophisticated (22) |
| Simple (110) | Useful (66) | Robust (36) | Benefit (20) |
| Good (100) | Successfully (56) | Considerable (36) | Simpler (19) |

**Table 3.** Most frequent bigrams with positive polarity

| | | | |
|---|---|---|---|
| improvement in | success of | more efficient | most successful |
| good performance | can improve | very successful | most notable |
| good result | more accurate | best score | well known |
| development of | most important | widely used | effective at |
| high quality | achieve impressive | quite accurate | increase over |

### 4.3 Negative polarity words (NPW)

Our next feature is obtained by using a collection of negative words. The set of negative words which are often encountered in scientific literature is relatively small in number. Owing to peer relations, criticisms of scientific papers are often hedged and implicit. Thus, we check the citation sentence to find any word which belongs to this collection of negatively polar words. Table 4 illustrates the negative words which usually occur in citation texts.

**Table 4.** Most frequent negative polarity words in citations

| | | | |
|---|---|---|---|
| However (125) | Unlike (26) | Worse (8) | Unrealistic (2) |
| While (119) | Restrict (14) | Unfortunately (7) | Insufficient (1) |
| Although (68) | Lack (13) | Complicated (5) | Inability (1) |
| Low (45) | Poor (10) | Daunting (4) | Lack of (12) |
| Without (40) | Unexplored (9) | Degrade (3) | Not well (2) |

| Difficult (36) | Little (8) | Burden (3) | Not able to (1) |
| --- | --- | --- | --- |

### 4.4 Presence of specific Part-Of-Speech tags (POS)

The output of the POS tagger is analyzed to check for the presence of specific tags like JJ, JJR, JJS, JJT (various forms of adjective), RB, RBR, RBT, RN, RT (forms of adverbs) and FW (foreign words). We also checked the occurrence of adverbs followed by adjectives (for example RB_JJ tag) as the presence of adverb along with adjective usually reflects subjectivity in sentence polarity.
*e.g.: simpler/ JJR, well/RB, etc.*
Here, *simpler* and *well* are two subjective words which are tagged as JJR and RB respectively.

### 4.5 Presence of specific Dependency tags (DEP)

While obtaining dependency output, we check for the presence of tags *advmod* (adverb modifier), *acomp* (adverbial complement) and *amod* (adjectival modifier) in the sentence. These tags are also indicators of subjectivity in sentence.
*e.g.: simpler approach, well known, etc.*

Here, *amod* (approach, simpler) and *advmod* (known, well) captures the polarity of citation. Similarly, *acomp* functions like an object of the verb and *amod* is any adjectival phrase that modifies the meaning of the noun phrase (NP). These relations are most frequent in sentences where polar sentiments are present.

### 4.6 Self Citation (SC)

We also check the presence of self citation in the citation sentence. This can be checked by verifying if the citing (source) paper refers to itself. We checked that there were no self citations in our dataset. When we constructed a graph representing citations, we found that it contained no self-loop. So we did not include this feature for classification purposes.

### 4.7 Opinion Lexicons (OL1 and OL2)

This feature is identified from a list of positive and negative opinion or sentiment words. The list was developed by Liu et al. [18] for comparing opinions on the web. We have used this list for identifying any sentiment words in the text. We have also used Vender Sentiment, which is another sentiment word list for identifying the sentiment words and determining their polarity. Both these lists have been split into positive and negative collections and then four features were introduced – the number of matches to each list - to train the classifier.

# 5    CLASSIFICATION

We used the machine learning software WEKA[2] [19]. We combined the above features to form a feature set and used the J48 classifier to generate a pruned C4.5 Decision Tree for three-way classification of the citation instances – *positive, negative* and *neutral*. The C4.5 algorithm generates a classification-decision tree for the given dataset by recursive partitioning of the data. It uses depth-first strategy and makes the selection based on highest information gain.

Individually, none of the features was able to detect positive or negative instances in citation. This was due to the large number of neutral instances present in the system and biasness of such neutral instances. We performed feature analysis by removing one feature at a time to determine if any feature was more important than the other. We also checked by adding one feature at a time.

The classification confidence score from WEKA and the number of matches to our citation specific lexicon were used to develop a post-processing algorithm. We added extra weight to the frequency of matches to the lexicon list. If the difference between frequency of positive and negative polarity words was more than $t_1$, we immediately assigned the instance as positive citation. If the number of negative polarity words was more than $n_1$, we assigned it as negative. Next we considered the confidence score of our WEKA classification. If it is more than $s_1$, we use the WEKA classification. Otherwise we use the polarity matches again to determine the polarity. The thresholds for this step are $t_2$ and $n_2$ respectively. This algorithm helped us to improve the accuracy of our result. Focusing on the best results obtained for different values of $t_1$ and $n_1$, ranging from one to five, and s1, ranging from 0.5 to 1.0, we settled for $t_1 = 3$, $n_1 = 2$, and $s_1 = 0.8$. Similarly best results were obtained by setting $t_2 = 2$ and $s_2 = 1$. Note that traditional accuracy measures are often not a good metric when the classes are imbalanced and/or cost of misclassification varies dramatically between the two classes.

## 5.1    Feature Analysis

**Table 5.** Impact of each feature calculated by eliminating one at a time

| Feature eliminated | Number of correct classifications | Number of incorrect classifications | Accuracy |
|---|---|---|---|
| SWN Lexicon | 1740 | 260 | 0.87 |
| Citation specific lexicons | 1740 | 260 | 0.87 |
| Part of speech tags | 1731 | 269 | 0.8655 |
| Dependency tags | 1732 | 268 | 0.866 |
| Opinion Lexicon 1 | 1740 | 260 | 0.87 |
| Opinion Lexicon 2 | 1722 | 278 | 0.861 |

---

[2] http://www.cs.waikato.ac.nz/ml/weka/downloading.html

Table 6. Impact of adding each feature iteratively to the last

| Feature added | Number of correct classifications | Number of incorrect classifications | Accuracy |
|---|---|---|---|
| SWN Lexicon | 1740 | 260 | 0.87 |
| Citation specific lexicons | 1740 | 260 | 0.87 |
| Part of speech tags | 1746 | 254 | 0.873 |
| Dependency tags | 1744 | 256 | 0.872 |
| Opinion Lexicon 1 | 1722 | 278 | 0.861 |
| Opinion Lexicon 2 | 1736 | 264 | 0.868 |

## 5.2 Algorithm

Table 7. Algorithm to classify the citation instances

```
ALGORITHM : Program Classification
    begin
        L1 = citation specific lexicon list
        score = classification confidence score from WEKA
        C = Class assigned by WEKA
        posmatch = number of matches to positive polarity words
        negmatch = number of matches to negative polarity words
        if posmatch-negmatch > t1, class = "positive"
        else if negmatch > n1, class = "negative"
        else if score > s1, class = C
        else if posmatch – negmatch > t2, class = "positive"
        else if negmatch > n2, class = "negative"
        else class = "neutral"
    end
```

## 5.3 Results

Table 8 shows the confusion matrix for the polarity classification. The precision, recall and f-measure of the supervised and baseline systems are compared in Table 9.

Table 8. Confusion Matrix for the classification result

|  | Positive | Neutral | Negative |
|---|---|---|---|
| Positive | 33 | 159 | 2 |
| Neutral | 27 | 1704 | 8 |
| Negative | 3 | 51 | 13 |

Table 9. Precision, Recall and F-measure of supervised system and the baseline

|  | Precision | Recall | F-measure |
|---|---|---|---|
| Supervised system | | | |
| Class Positive | 0.524 | 0.17 | 0.257 |
| Class Neutral | 0.889 | 0.968 | 0.927 |
| Class Negative | 0.545 | 0.179 | 0.27 |
| Baseline System | | | |
| Class Positive | 0 | 0 | 0 |
| Class Neutral | 0.864 | 1 | 0.927 |
| Class Negative | 0 | 0 | 0 |

The baseline model considered all the instances to be of neutral polarity. So we can see that our supervised system shows improvement over the baseline model. However, the learning algorithm was slightly biased towards neutral classification which is evident from the confusion matrix. Most of the errors are due to positive and negative citations being identified as neutral.

In future works, we will need to fine tune our classification features so that the system can identify positive and negative citations more efficiently. Also using a larger dataset to train the system would eliminate the bias towards neutral classification of polarity.

## 6   RANKING ALGORITHM

For scientific literature, we generally use the H-index to find out the impact of an author. However, H-index is only a quantitative measurement. Also, it targets an author in particular. Our approach captures the importance of each paper based on the number and the opinion of the citing paper. We also capture both quantitative and qualitative factors in our index. The ranking index should help in identifying the paper which has not only been cited more often but also in a positive sense. By taking into account the criticism of the citing papers, we aim to make the ranking system more efficient and feasible.

### 6.1   Corpus

The dataset contains the following information:

**Citation sentence.**
This is the sentence in a research paper where the paper refers to one or more scientific papers in order to explain, state, praise or criticize the claims made in the referred paper(s). While the first two cases would be of neutral polarity, the last two cases are of positive and negative polarity respectively.
*e.g.- Dasgupta and Ng (2007) improves over (Creutz, 2003) by suggesting a simpler approach.*

**Source.**
This is the paper id of the source or citing paper. This paper quotes or borrows some idea or concept explained in the cited paper. In the previously mentioned example, the source paper id is '*W09-0805*', where we find the citation sentence.

**Target.**
This is typically the paper id of the cited paper. If we represent each citation instance using an edge of the graph, and the papers themselves as nodes of the graph, then we can find a directed edge from the source or citing paper to the target or cited paper. In the previously mentioned citation sentence, the target or cited paper id is '*N07-1020*'. This is the paper id for the paper by Dasgupta and Ng (2007).

**Polarity (calculated by our system) .**
The polarity is calculated by our sentiment classification system in the first place. We have considered only three types of polarity – positive, negative and neutral.

### 6.2    Naïve Algorithm

The naïve algorithm is the standard baseline which counts the number of times a particular paper is cited by other papers.

**Algorithm.**

Table 10. Naïve Algorithm to find the ranking of papers

| ALGORITHM |
|---|
| begin |
|     total_papers = Total number of papers in the collection |
|     total_instance = Total number of instances |
|     for (num=0; num<total_papers; num++) |
|         paper.count = 0; |
|     end for |
|     for each instance i |
|         for all the papers in collection |
|             if paper.id == target_paper.id |
|                 paper.count +=1; |
|                 break; |
|             end if |
|         end for |
|     end for |
|     sort all the papers by count to obtain ranked index |
| end |

**Ranking.**

The ranked list of papers was divided into buckets. Each bucket comprised 20% of the total number of cited papers. In our test set, there were 40 unique target (cited) papers for 2000 citation instances. So, we divided the ranked list into 5 buckets with each bucket consisting of 8 papers sorted by rank.

### 6.3 Proposed Algorithm (M-index)

For scientific literature, we use the H-index to find out the impact of an author. However, H-index is only a quantitative measurement. We aim to capture both quantitative and qualitative factors in our index. So, we evaluated the impact of the paper based on two factors – the number of citations that the paper received (quantitative) and the polarity of the citation (qualitative). For the proposed algorithm, we have used three different kinds of scores – the reliability score, the polarity score and the M-Index Score.

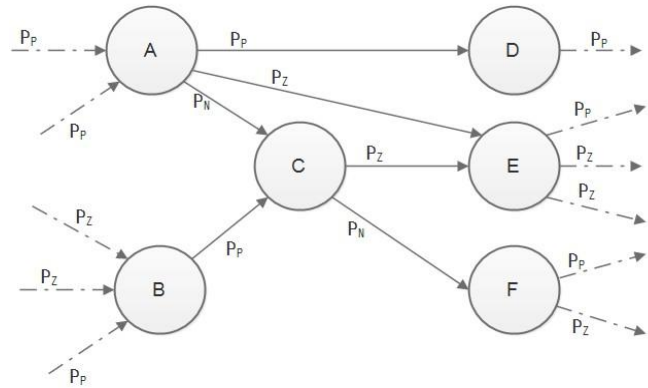

**Fig. 1.** Graphical Representation of Citations

We denote the citation dataset in the form of a directed graph $G = <V, E>$ where V is a set of all nodes and E is the set of directed edges over those nodes. Each node represents a research paper (cited or citing or both) and each outgoing edge represents a citation instance from the source (citing) node to the target (cited) node. The edges are marked with polarity scores of the instance.

**Polarity Score.**

Using our previous classification system, we judged the polarity of each citation instance. Polarity Score is denoted by PS(e) for any edge or instance $e \in E$. In Figure 1, we have used $P_P$, $P_Z$, and $P_N$ to denote positive, neutral and negative polarity respectively. We have assigned a polarity of +1, 0.5 and -0.5 to papers of positive, neutral and negative polarity respectively for each citation instance. Therefore, $P_P = 1$, $P_Z$

= 0.5 and $P_N$ = -0.5. In Figure 1, node C has two incoming edges with polarities $P_N$ and $P_P$ while node B has three incoming edges $P_Z$, $P_Z$ and $P_P$. We have tuned our system to reward positive citations by a larger weight. The weights of neutral and negative citations are kept the same. Neutral citations have been assigned with a positive score instead of zero to reward non-negative instances of citations. The polarity score is relevant only for the target or cited paper. It shows if the target paper has been referred subjectively and the polarity of that reference.

**Reliability Score.**

For each paper, we can find out the extent to which we can rely on the paper for citations. If a paper has been cited negatively by most other papers, then we can assume that the paper lacks reliability. So we assign a score of +2, +1 and -1 to denote that the paper is *very reliable*, *fairly reliable* and *not reliable* respectively. In some cases, we may not be able to find the reliability of the source paper. This is due to the fact that the source paper may not appear as target in any citation instance. In such a situation, we will consider the source paper as *fairly reliable*, i.e., assign it a reliability score of +1. The reliability score is applicable only for the source or citing paper because we are concerned with the capacity of judgment of the citing paper. The reliability score, denoted by R (n) for a node n ∈ V, is calculated by the summation of the polarity scores of all the incoming edges. In Figure 1, the reliability score of node C can be calculated by finding the sum $P_Z, P_Z$ and $P_P$.

$$R(n) = \sum_{e \in E_n} PS(e) \quad (1)$$

Where $E_n$ is the number of incoming edges for node n.

The reliability score, R(n), thus obtained, is normalized to +2, +1 and -1 respectively for different ranges (as explained in the algorithm of Table 11). The normalized reliability score is denoted by R'(n).

**M-Index Score.**

The Instance Score, denoted by I(e) for each edge e ∈ E, is defined as the weighted score of each citation instance. This score was calculated by taking into account the reliability of the source or citing paper (or node) and the sentiment polarity of each reference. For a given node or paper n ∈ V, we calculated MIS(n) as the sum of instance scores for all the incoming edges of node n.

$$I(e) = \left(PS(e) * R'^{(n)}\right) \quad (2)$$

$$MIS(n) = \sum_e I(e) \quad (3)$$

**Algorithm.**

Table 11. Algorithm to determine ranking of papers by m-index

| ALGORITHM |
|---|
| Begin |
|   total_papers = total number of papers in the collection |
|   total_instances = total number of instances in the collection |
|   for (num = 0; num < total_papers; num++) |
|     paper.relscore = 0 |
|   for each instance i |
|     for all papers |
|       if the paper.id = target.id |
|         if ( polarity > 0 ) paper.relscore += pp |
|         if ( polarity < 0) paper.relscore -= np |
|         if ( polarity == 0 ) paper.relscore += zp |
|       end if |
|     end for |
|   end for |
|   for (num=0; num< total_papers; num++) |
|     paper.mscore = 0; |
|     if (paper.relscore > 1)  paper.relscore = 2 |
|     else if (paper.relscore < 0)  paper.relscore = -1 |
|     else  paper_relscore = 1 |
|   end for |
|   for each citation instance i |
|     score1 = sourcepaper.relscore |
|     score2 = targetpaper.polarityscore |
|     instance_score = score1 * score2 |
|     targetpaper.mscore +=  instance_score |
|   end for |
|   sort cited papers by m-score to get the ranking index |
| end |

**Ranking.**

A total of 2000 citation instances had 40 unique cited papers. These papers were ranked by m-score. The ranked list was divided into 5 buckets, each bucket containing 8 papers.

## 7    RESULT ANALYSIS

After ranking the papers based on naïve method and the modified algorithm, we divided them into 5 buckets and tried to evaluate the impact of the modified algorithm on each bucket. It did not present much variation in the overall ranking which was understandable owing to the limitations of the dataset. Only 3 papers out of 40 showed variations in the ranking and even then, the buckets were not altered.

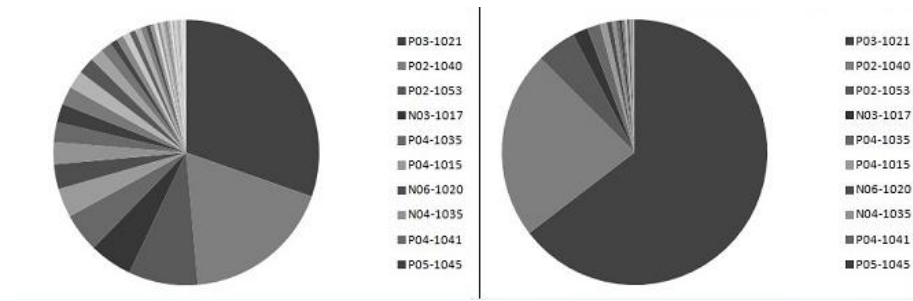

**Fig. 2.** Score distribution of ranked papers (naïve score on the left and m-score on the right)

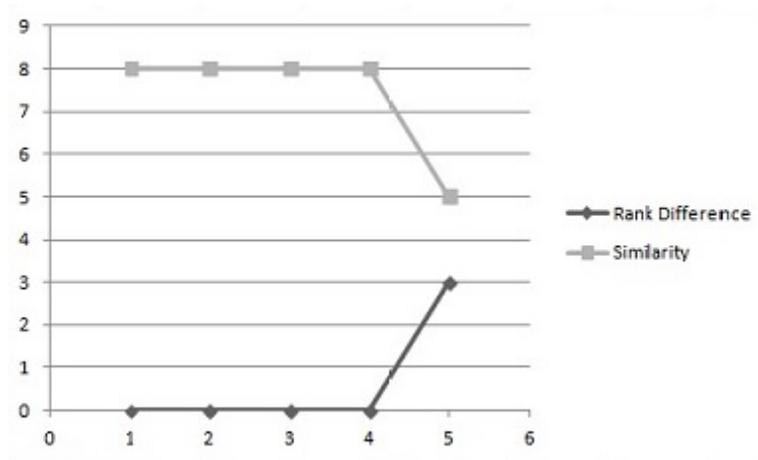

**Fig. 3.** Rank difference and similarities between the 5 buckets

The limited nature (imbalance and small size) of the corpus was one of the primary reasons as to why there was not much variation between the two ranked lists. If we look at the pie-chart of Figure 2, we can notice that the top-ranked papers were clearly cited much more than the remaining ones. There was also a prominent gap in the number of citations for each paper. This resulted in fewer changes in ranking between the two methods.

In future works, we aim to concentrate on preparing a larger corpus which will contain a larger proportion of subjective citations. This would help in reducing the bias of our sentiment analysis algorithm towards neutral classification. Qualitative factors would also have a higher impact in the overall ranking.

## 8      CONCLUSION AND FUTURE WORK

In this paper, we focused on two aspects of citations, automatic detection of citation sentiment and ranking of scientific paper using a newly proposed index. First, we classified the polarity of citations using a statistical classifier C4.5 which made use of various sentence-based and linguistic features to generate decision trees. Our system achieved fairly accurate results with 87.5% accuracy.

Secondly, we proposed a new index, M-index, which takes into account the reliability of the citing paper and the type of polarity involved between the citing and cited paper. This index focuses mainly on a particular paper, unlike H-index, which is more author-specific. The ranked list of the cited papers was obtained using the new index. A similar ranked list was obtained using the naïve method which maintained a simple count of the number of times a paper was cited. We analyzed the impact of this new index by comparing the two ranked lists. Although the ranks did not show too much variation, yet the impact should be greater with a larger corpus.

For future work, we are working on a second corpus. This corpus is based on the ACL Anthology corpus[3] which has been annotated to take the dominant sentiment in the entire citation context into account. M-index based ranking uses both quantitative and qualitative information and its impact could be better analyzed by the larger corpus.

---

[3] http://clair.eecs.umich.edu/aan/index.php